\documentstyle[12pt]{article}

\author{Vadim L. Vereschagin\thanks{Supported by ISF Grant RK 2000}\\
Irkutsk Computing Centre, P.O.Box 1233, Irkutsk\\
664033, RUSSIA}
\title{NONLINEAR QUASICLASSICS AND THE PAINLEV\'E EQUATIONS
}
\input tcilatex

\QQQ{Language}{
American English
}

\begin{document}

\maketitle
{\bf 1. INTRODUCTION}

A century-old history of calculation of asymptotics for solutions to
Painlev\'e equations (usually denoted P$_j,$ j=1,2,...,6) as their variable $%
x$ tends to infinity was started by pioneer works by Painlev\'e, Gambier and
Boutroux \cite{PGB}. In 1980-1981 papers by Jimbo, Miwa and Flashka, Newell 
\cite{JMFN} initiated the so-called Isomonodromy Deformation Method (IDM)
which afterwards has given remarkable progress in this direction for P$_1,$ P%
$_2,$ and P$_3.$ Successful attempts to apply the Whitham-like method
connected with use of different-scaled variables were accomplished in papers 
\cite{KJIK}. This way yields leading term of the asymptotic series in form
of ''modulated'' elliptic function (Weierstrass $\wp $-function for P$_1$ or
Jacobi sine for P$_2$) such that parameters determining it depend on $x$ in
force of certain algebraic or differential equation. At last, the Whitham
approach in its ''classical'' interpretation was applied to P$_1$ by
Novikov, Dubrovin, Moore, Krichever \cite{NDMK}. These authors obtained
certain ODE usually referred to as ''Whitham ''or ''modulation'' equations.
In paper \cite{VER} solutions of Discrete Painlev\'e-1 eq. were investigated
from this point of view. The main goal of this paper is to generalize ideas
of the latter method and expand them to the remaining Painlev\'e eqs.

\medskip 

{\bf 2. IDEOLOGY OF THE METHOD}

Painlev\'e eqs are known to be integrable nonlinear ODE's. Probably, the
most explicit illustration of this fact is existence of commutative matrix
representation

\begin{equation}
\label{rep}\partial _\lambda L_j-\partial _xA_j+[L_j,A_j]=o,\ j=1,2,...,6, 
\end{equation}
where $\partial _x=\frac \partial {\partial x},$ $L_j=L_j(u,u^{\prime
},x,\lambda ),\ A_j=A_j(u,u^{\prime },x,\lambda )$ are 2$\times $2 matrices
rationally depending on spectral parameter $\lambda $ so that j-th
Painlev\'e eq. $u^{\prime \prime }-P_j(u,u^{\prime },x)=0$ is equivalent to (%
\ref{rep}). The matrices $L_j,\ A_j$ are listed, for example, in \cite{JMFN}.

Introduce now variable $X$ and change all variables $x$ explicitly entering 
$L_j,\ A_j$ to $X$ : $L_j=L_j(u,u^{\prime },X,\lambda ),\
A_j=A_j(u,u^{\prime },X,\lambda )$.

\underline{Proposition 1.} Let $\epsilon $ be some real positive number.
Then system

\begin{equation}
\label{epr}\epsilon \partial _\lambda L_j-\partial _xA_j+[L_j,A_j]=0 
\end{equation}

is equivalent to system

\begin{equation}
\label{epp} 
\begin{array}{c}
\partial _xX=\epsilon \\ 
u^{\prime \prime }-P_j(u,u^{\prime },X)=o 
\end{array}
\end{equation}

\underline{Proposition 2.} Solution of eq. (\ref{epr}) while $\epsilon =0$
and $X=const$ is

\begin{equation}
\label{sol0}u_0(x)=f_j(\tau +\Phi ;\overrightarrow{a}) 
\end{equation}
where $\tau =xU;\ f_j$ are periodic functions which can be represented in
terms of Weierstrass or Jacobi elliptic functions (see \cite{BE}) for any of
the six Painlev\'e eqs. $U=U(X),\ \Phi =\Phi (X);\overrightarrow{a}(X)$ are
parameters determining the elliptic function $f_j$.

Suppose now $\epsilon $ to be small and positive. We shall look for
solutions of (\ref{epp}) as formal series in $\epsilon :$

\begin{equation}
\label{ser}u(x)=u_0(x)+\epsilon u_1(x)+... 
\end{equation}
so that parameters determining the elliptic functions $u_0=f_j$ obey some
special ODE called ''Whitham'' or ''modulation'' system. Thus, we look for
the leading term of (\ref{ser}) in form $u_0(\tau ,X)=f_j\left( \frac{S(X)}%
\epsilon +\Phi (X);\overrightarrow{a}(X)\right) ,$ where $\partial _XS=U.$
The Whitham system can be easily derived from eq. (\ref{epr}) by simple
change $\partial _x\rightarrow U\partial _\tau +\epsilon \partial _X$ and
averaging in $\tau :$

\begin{equation}
\label{whi}\partial _X\det A_j=\overline{a_{22}\partial _\lambda l_{11}}+ 
\overline{a_{11}\partial _\lambda l_{22}}-\overline{a_{12}\partial _\lambda
l_{21}}-\overline{a_{21}\partial _\lambda l_{12}} 
\end{equation}
where $A_j=(a_{mn}),\ L_j=(l_{mn}),\ m,n=1,2;$ bar means averaging over
period of the elliptic function (\ref{sol0}).

\underline{Proposition 3. }There exists only unique coefficient in the
polynomial det$A_j$ whose dynamics in $X$ is non-trivial. Denote this
coefficient $F_j.$ Therefore the Whitham system can be written out as a
single first-order ODE on $F_j$.

\underline{Corollary.} The simplest way to find elliptic ansatz is to solve
equations

\begin{equation}
\label{con}F_j=const_{1,}\ X=const_2 
\end{equation}

\underline{Proposition 4.} The Whitham system induces formal condition

$$
u_0^{\prime \prime }-P_j(u_0,u_0^{\prime },x)=O(\epsilon ) 
\TeXButton{footnote}
{\footnote{This condition makes rigorous sense only for smooth function u. 
However, the method describes singular solutions as well.}}
$$

Obviously, the scaling transformation $x\rightarrow \epsilon x$ induces $%
\partial _xX=\epsilon \rightarrow \partial _xX=1$ in (\ref{epp}), which
provides the following proposition:

\underline{Proposition 5.} The function $u_0$ specified by (\ref{sol0}) and (%
\ref{whi}) yields leading term of asymptotic series for solution of
appropriate Painlev\'e equation as $x$ tends to infinity.

\medskip

{\bf 3. CONCRETE CALCULATIONS.}

P$_1:\quad u^{\prime \prime }-3u^2-x=0$

$\det A_1=16\lambda ^3+4X\lambda -F_1,\ where\ F_1=(u^{\prime
})^2-2u^3-2Xu;\ \partial _X\det A_j=2\overline{u}+4\lambda \Rightarrow $

$\partial $$_XF_1=-2\overline{u}=-2\eta /\omega =2e_1+2(e_3-e_1)\frac EK,$
where $K,\ E$ are complete elliptic integrals:

$K=\int\limits_0^1\frac{dz}{\sqrt{(1-z^2)(1-k^2z^2)}},\quad
E=\int\limits_0^1\sqrt{\frac{1-k^2z^2}{1-z^2}}dz,\quad k^2=\frac{e_2-e_3}{%
e_1-e_3};$ $e_1,\ e_{2,}\ e_3$ are roots of the polynomial $%
R_3(z)=4z^3-g_2z-g_3$, $g_2=-X,\ g_3=-\frac 14F_1.$

$u_0=2\wp (x+\Phi ;g_2,g_3).$

\medskip\ 

$P_2:\quad u^{\prime \prime }-2u^3-ux=0$%
\footnote{Case with additional constant can be investigated in the same way}

$\det A_2=16\lambda ^4+8X\lambda ^2+X^2-4F_2,$ where $F_2=\left( u^{\prime
}\right) ^2-u^4-Xu^2;$

$\partial _X\det A_2=8\lambda ^2+2X+4\overline{u^2}\Rightarrow \partial
_XF_2=-\overline{u^2}=2\frac \eta \omega -e_m=X\left( \frac 12-\frac
1{2-k^2}\frac EK\right) ,$ where $k^2=\frac X{2F_2}(X\mp \sqrt{X^2-4F_2})-1;$

$u_0=\sqrt{-\frac 12(x\mp \sqrt{x^2-4F_2})}sn\left( \frac x{\sqrt{2}}\sqrt{%
-x\pm \sqrt{x^2-4F_2}}+\Phi ;k\right) ,\quad sn$ is Jacobi sine.

\medskip\ 

$P_3:\quad u^{\prime \prime }+\frac 1xu^{\prime }+\sin u=0$

$\lambda ^4\det A_3=\frac{\lambda ^4X^4}{256}+\frac{\lambda ^2X^2}8F_3+1,$
where $F_3=\frac{\left( u^{\prime }\right) ^2}2-\cos u;$

$\partial _XF_3=-\frac 2X(F_3+\overline{\cos u})=\frac 1{Xk^2}\left(
k^2-1+2\frac EK\right) ,\ k^2=\left(-F_3\pm \sqrt{F_3^2-1}\right)^2;$

$u_0=-2i\log \left( ksn\left( \frac{ix}{2k}+\Phi ;k\right) \right) $

\medskip\ 

$P_4:\quad u^{\prime \prime }=\frac{\left( u^{\prime }\right) ^2}{2u}+\frac{%
3u^3}2+4xu^2+u(a-4d+2x^2)-\frac{a^2}{4u},$
where $a,d$ are arbitrary parameters.
$\det A_4=-\frac{\lambda ^6}4-X\lambda ^4-(d-a+X^2)\lambda ^2-F_4-\frac{d^2}{%
\lambda ^2},$

$F_4=\frac{\left( u^{\prime }\right) ^2}{4u}-\frac{u^3}4-Xu^2-u\left(
X^2+\frac a2-2d\right) +X(2d-a)-\frac{a^2}{4u},$

$\partial _X\det A_4=-\lambda ^4-2X\lambda ^2+\overline{u^2}+2X\overline{u}%
+a-2d\Rightarrow \partial _XF_4=-\overline{u^2}-2X\overline{u}-a+2d$
and $u_0=\pm U[\zeta (Ux+\Phi _1)-\zeta (Ux+\Phi _2)] \pm U\zeta (\Phi _2-
\Phi _1)-X,$ parameters $U,\ \Phi _2-\Phi _1$ can be
derived from (\ref{con}). The Whitham eq. looks too large to be written out
explicitly.

\medskip\ 

$P_5:\quad u^{\prime \prime }=\frac{3u-1}{2u(u-1)}\left( u^{\prime }\right)
^2-\frac{u^{\prime }}x+\mu \frac ux-\mu ^2\frac{u(u+1)}{2(u-1)},\ \mu $ is a
parameter;

$\det A_5=-4X^2-\frac{F_5}{\lambda (\lambda -1)},\quad F_5=\frac{X^2}{%
4u(u-1)^2}\left[ \left( u^{\prime }\right) ^2-2uu^{\prime }(\mu -4)+u^2\mu
(\mu -8)\right] ;$

$\partial _X\det A_5=-8X+\frac{4\overline{z}}{\lambda (\lambda -1)},$ where $%
z=\frac{X(\mu u-u^{\prime })}{2(u-1)^2}\Rightarrow \partial _XF_5=-4 
\overline{z};$

$u_0=\frac{U^2x^2}{F_5}\left[ \wp (Ux+\Phi _1)+\wp (Ux+\Phi _2)\right] $

$+\frac{U^3x^4(\mu -4)}{F_5(3F_5-U^2x^2)}\left[ \zeta (Ux+\Phi _1)-\zeta
(Ux+\Phi _2)\right] +V,$

parameters $U,\ \Phi _1-\Phi _2,\ V$ can be derived from (\ref{con}).

\medskip\ 

$P_6:\quad u^{\prime \prime }=\frac 12\left( \frac 1u+\frac 1{u-1}+\frac
1{u-x}\right) \left( u^{\prime }\right) ^2-\left( \frac 1x+\frac
1{x-1}+\frac 1{u-x}\right) u^{\prime }$

$+\frac{u(u-1)(u-x)}{x^2(x-1)^2}\left[ \alpha +\beta \frac x{u^2}+\gamma 
\frac{x-1}{(u-1)^2}+\delta \frac{x(x-1)}{(u-x)^2}\right] ,$

$\alpha ,\ \beta ,\ \gamma ,\ \delta $ are parameters;

$\lambda \left[ \left( \lambda -1\right) (\lambda -x)\right] ^2\det
A_6=k_1k_2\lambda ^3+F_6\lambda ^2+S_1\lambda +S_0;$

One can determine $F_6$ and the elliptic ansatz from the following
constraint:

$$
\left( u^{\prime }\right) ^2x^2(x-1)^2+u^{\prime
}a_0+(k_1-k_2)^2u^4+a_3u^3+a_2u^2+a_1u+\theta _0^2x^2=0, 
$$
where:

$\frac 12a_0=u^2x(x-1)(k_1+k_2)+ux\left[ x^2(\theta _0+\theta _1)+x(\theta
_2-\theta _1)-\theta _0-\theta _2\right] -\theta _0x^3+\theta _0x^2,$

$\frac 12a_3=4k_2(k_1-k_2)(x+1)+(3k_2-k_1)[x(\theta _0+\theta _1)+\theta
_0+\theta _2]-2F_6,$

$a_2=x^2[(\theta _0+\theta _1)(\theta _0+\theta
_1+4k_1-4k_2)+4k_2(2k_2-k_1)]+2x[10k_2^2-6k_1k_2-2k_1^2-4k_1\theta _0$

\quad $+\theta _2\theta _1+2F_6]+4k_2(2k_2-2k_1)+(\theta _0+\theta
_2)(\theta _0+\theta _2+4k_1-4k_2)+4F_6,$

$\frac 12a_1=x^2[(\theta _0+\theta _1)(2k_2-2k_1-\theta
_0)+2k_2(k_1-2k_2)]+x[(\theta _0+\theta _2)(2k_2-2k_1-\theta _0)+$

$+2k_2(k_1-2k_2)]-2F_6;$

$k_1+k_2=-(\theta _0+\theta _1+\theta _2),\ k_1-k_2=\theta _\infty ,\ \alpha
=\frac 12(\theta _\infty -1)^2,\ \beta =-\frac 12\theta _0^2,\ $

$\gamma =\frac 12\theta _1^2,\ \delta =\frac 12\left( 1+\theta _2^2\right) ;$

$\partial _XF_6=-2k_1k_2+z_2(k_1-k_2)+\theta _2(z_2-k_2)+\frac{(z_0+\theta
_0)z_0(u-x)}{u(x-1)}-\frac{(z_1+\theta _1)z_1(u-x)}{x(u-1)},\ where$

$z_0=\frac u{x\theta _\infty }\{u(u-1)(u-x)z^2+\left[ \theta _1(u-x)+x\theta
_2(u-1)-2k_2(u-1)(u-x)\right] z+$

$+k_2^2(u-x-1)-k_2(\theta _1+x\theta _2)\},$

$z_1=-\frac{u-x}{(x-1)\theta _\infty }\{u(u-1)(u-x)z^2+$

$\left[ (\theta _1+\theta _\infty )(u-x)+x\theta
_2(u-1)-2k_2(u-1)(u-x)\right] z+k_2^2(u-x)-k_2(\theta _1+x\theta
_2)-k_1k_2\},$

$z_2=\frac{u-x}{x(x-1)\theta _\infty }\{u(u-1)(u-x)z^2+$

$\left[ \theta _1(u-x)+x(\theta _2+\theta _\infty
)(u-1)-2k_2(u-1)(u-x)\right] z+k_2^2(u-1)-k_2(\theta _1+x\theta
_2)-xk_1k_2\},$

$z=\frac{u^{\prime }x(x-1)-\theta _0(u-1)(u-x)-\theta _1u(u-x)-\theta
_2u(u-1)}{2u(u-1)(u-x)}$

\end{document}